\documentclass[12pt]{iopart}
\usepackage{graphicx}

\newcommand{\ts}[2]{{#1}_{\textnormal{#2}}} 
\newcommand{\tsc}[2]{{#1}_{\textsc{#2}}} 
\newcommand{\tn}[1]{\textnormal{#1}}
\newcommand{\tc}[1]{\textsc{#1}}

\newcommand{\be}{\begin{equation}}
\newcommand{\ee}{\end{equation}}

\newcommand{\ket}[1]{\left| #1 \right\rangle}  
\newcommand{\bra}[1]{\left\langle #1 \right|}  
\newcommand{\braket}[2]{\left\langle #1 |#2\right\rangle} 

\newcommand\ginzton{E. L. Ginzton Laboratory,
        Stanford University,
        Stanford, California 94305, USA}
\newcommand\uTokyo{Institute of Industrial Science, University of Tokyo,
        4-6-1 Komaba, Meguro-ku,
        Tokyo 153-8505, Japan}
\newcommand\panasonic{Nanotechnology Research Laboratory,
        Advanced Technology Research Laboratories, Panasonic Corporation,
        3-4 Hikaridai, Seika-cho, Soraku-gun,
        Kyoto 619-0237, Japan}
\newcommand\NII{National Institute of Informatics,
        Hitotsubashi 2-1-2, Chiyoda-ku,
        Tokyo 101-8403, Japan}
\newcommand\simonfraser{Department of Physics,
        Simon Fraser University,
        Burnaby, British Columbia V5A IS6, Canada}
\newcommand\HRL{HRL Laboratories, LLC,
        3011 Malibu Canyon Road,
        Malibu, California 90265, USA}

\begin{document}
\title{Quantum Hall Charge Sensor for Single-Donor Nuclear Spin Detection in Silicon}
\date\today

\author{D Sleiter$^1$, N Y Kim$^{1,2}$, K Nozawa$^3$, T D Ladd$^{1,4}$\footnote{Present address: \HRL}, M L W Thewalt$^5$ and Y Yamamoto$^{1,4}$}
\address{$^1$ \ginzton}
\address{$^2$ \uTokyo}
\address{$^3$ \panasonic}
\address{$^4$ \NII}
\address{$^5$ \simonfraser}
\ead{dsleiter@stanford.edu}

\begin{abstract}
We propose a novel optical and electrical hybrid scheme for the
measurement of nuclear spin qubits in silicon. By combining the
environmental insensitivity of the integer quantum Hall effect
with the optically distinguishable hyperfine states of
phosphorus impurities in silicon, our system can simultaneously offer
nuclear spin measurement and robustness against environmental
defects. $^{31}$P donor spins in isotopically purified
$^{28}$Si are often discussed as very promising
quantum memory qubits due to their extremely long decoherence
times, and our proposed device offers an effective
implementation for such a quantum memory system.
\end{abstract}

\submitto{\NJP}

\maketitle

\section{Introduction}

Semiconductor qubit implementations hold many advantages over
other qubit systems, such as long decoherence times and a
wealth of current industry experience. While solid state
systems are very complex and provide many challenges due to
many-body interactions, defects, and disorder, a number of experiments have shown
measurement and coherent control of spin qubits in
semiconductor devices~\cite{pettaDots05,vandersypenQD,hansonQD05,pressnature,duttGaAsQD,greilichQDensemble,FuGaAs,jiang}.

Nuclear spins in semiconductors have particularly long
decoherence times \cite{lmyai05}. The $^{31}$P donor nucleus in
Si has a spin relaxation time of thousands of
seconds~\cite{fg59}, and has recently been used to store
quantum coherence for several seconds~\cite{mtbslashal08}.
These long relaxation times motivated the seminal Kane proposal
for using the $^{31}$P nuclear spin in Si as a potential
semiconductor qubit~\cite{kane98}. However, individual
donor nuclear spin states have not yet successfully been
detected to date.  Nuclear spin states have only been measured in large
ensembles using either magnetic resonance techniques~\cite{feher56} or, more
recently, optical spectroscopy of $^{31}$P donor-related
transitions~\cite{thewalt}.

A few recent proposals outline how a single donor spin could
be measured either electrically~\cite{whaleyprop} or
optically~\cite{flsy04}. These proposals face a large number of
hurdles, beginning with the difficulty of isolating a single
donor.  We may compare these prior works by dividing each
proposal into two phases: a ``pump" phase, in which
spin-selective transitions are driven, and a ``detection"
phase, in which a scattering process reveals the result of the
pump phase.

Optical techniques excel in the pump phase due to the easily
distinguishable hyperfine-split optical transitions in
isotopically purified $^{28}$Si~\cite{thewalt}.  However,
optical detection is very challenging because of the extremely
inefficient radiative recombination due to the indirect bandgap in Si, requiring heroic efforts in cavity quantum electrodynamics to enhance the weak emission of
a single donor~\cite{flsy04}.   Electrical methods have
achieved great success in the detection component of the
measurement~\cite{jiang}, but microwave pumping introduces more
noise processes, quickly relaxing the measured spin~\cite{whaleyprop}. Microwave fields are also difficult to localize to a single device, an important consideration for future quantum computers.

Here, we propose a novel scheme that employs the advantages of
both optical and electrical measurement techniques in order to
overcome the difficulties of each. By combining the hyperfine
selectivity of optical pumping with the sensitivity of
electrical detection, our proposed measurement device can
perform deterministic quantum non-demolition projective
measurements of a single $^{31}$P donor nuclear spin in Si. One scheme of note that combines optical pumping with electrical detection together, the ``optical nuclear spin transistor", was previously mentioned in~\cite{thewalt}. We extend this scheme by using a quantum Hall bar device instead of a normal transistor to perform the electrical measurement. By employing the integer quantum Hall effect (IQHE) in a conductance plateau region, electrical noise due to
defects and background magnetic fields are suppressed.
Additionally, a quantum point contact (QPC) makes the
electrical device sensitive to only the small volume
surrounding the single $^{31}$P donor, isolating the desired
signal from noise due to other impurities and the electrical
contacts.

Besides offering a measurement technique which
overcomes difficulties of existing measurement proposals, the
IQHE may also introduce a possible method for performing
two-qubit gates between two donor nuclei~\cite{privmanQHQC}, where the extended-state edge channels can coherently couple two donors. In addition to the possibilities for gate and measurement operations, this scheme has many advantages in terms of controllability and integration. Many of the donor interaction parameters can be electrically controlled by the Hall bar device, such as the strength of the interaction between the edge channels and the donor. Furthermore, this device can be fabricated consistent with current CMOS fabrication techniques and is easily integrated with other electronics on the same chip, as opposed to other systems such as diamond.

In \sref{DescriptionSection} we explain the measurement scheme and device
structure, and then in \sref{DevicePhysicsSection} we elaborate on the
particular interactions and effects occurring within our system, discuss our simulation of the device physics, and explain how many of the interactions can be tuned by the device parameters.

\section{Description of Device and Measurement Scheme}
\label{DescriptionSection}

\begin{figure}[b]
\begin{center}
  \includegraphics[width=3.5in]{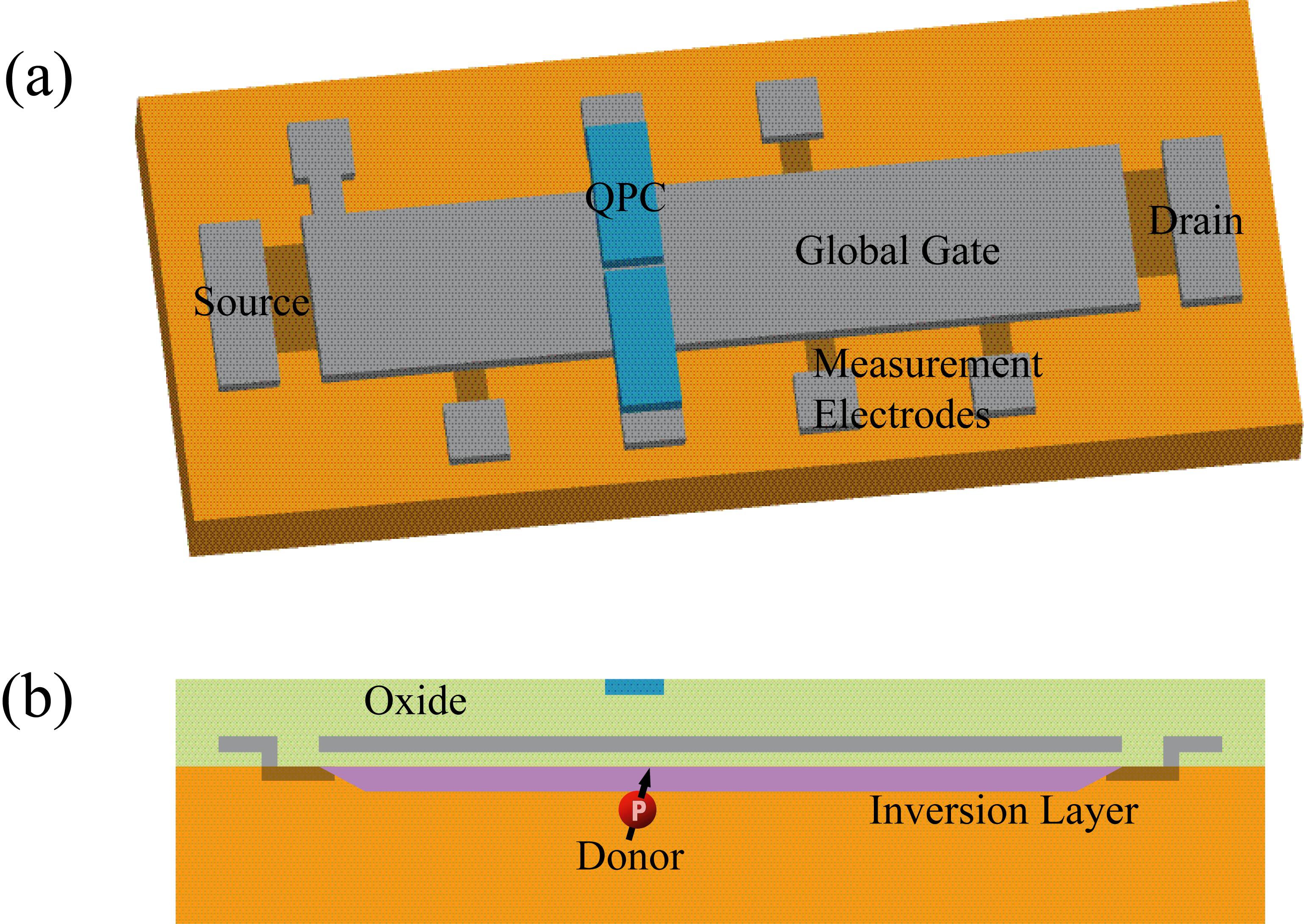} 
\end{center}
  \caption{A device schematic showing the important components. The device is comprised of a MOSFET Hall bar with a global gate to create and tune the inversion layer. Above the donor and global gate is the QPC gate (blue), separated from the rest of the device by an oxide layer (green). An aperture in the global gate (not shown) allows optical illumination in the proximity of the donor while preventing illumination of the source, drain, and measurement electrodes.}
  \label{deviceFig}
\end{figure}

Our device is composed of three main components (\fref{deviceFig}): a
basic MOSFET Hall bar device, a single P donor, and a QPC which surrounds the donor. The measurement also employs one or two external narrow linewidth continuous wave (CW) lasers which can be tuned to the set of neutral donor (D$^0$) to donor bound
exciton (D$^0$X) optical transitions. We will first describe
the system in an ideal case, where many of the complex
interactions within semiconductors are ignored (these effects
are discussed in section III). For now, we assume a perfect
$^{28}$Si crystal with only one $^{31}$P donor, we ignore the
effects of the oxide and electric field on the donor, and
likewise neglect spin-spin scattering between conduction
electrons and the donor electron.

\begin{figure}[b]
\begin{center}
\includegraphics[width=3.5in]{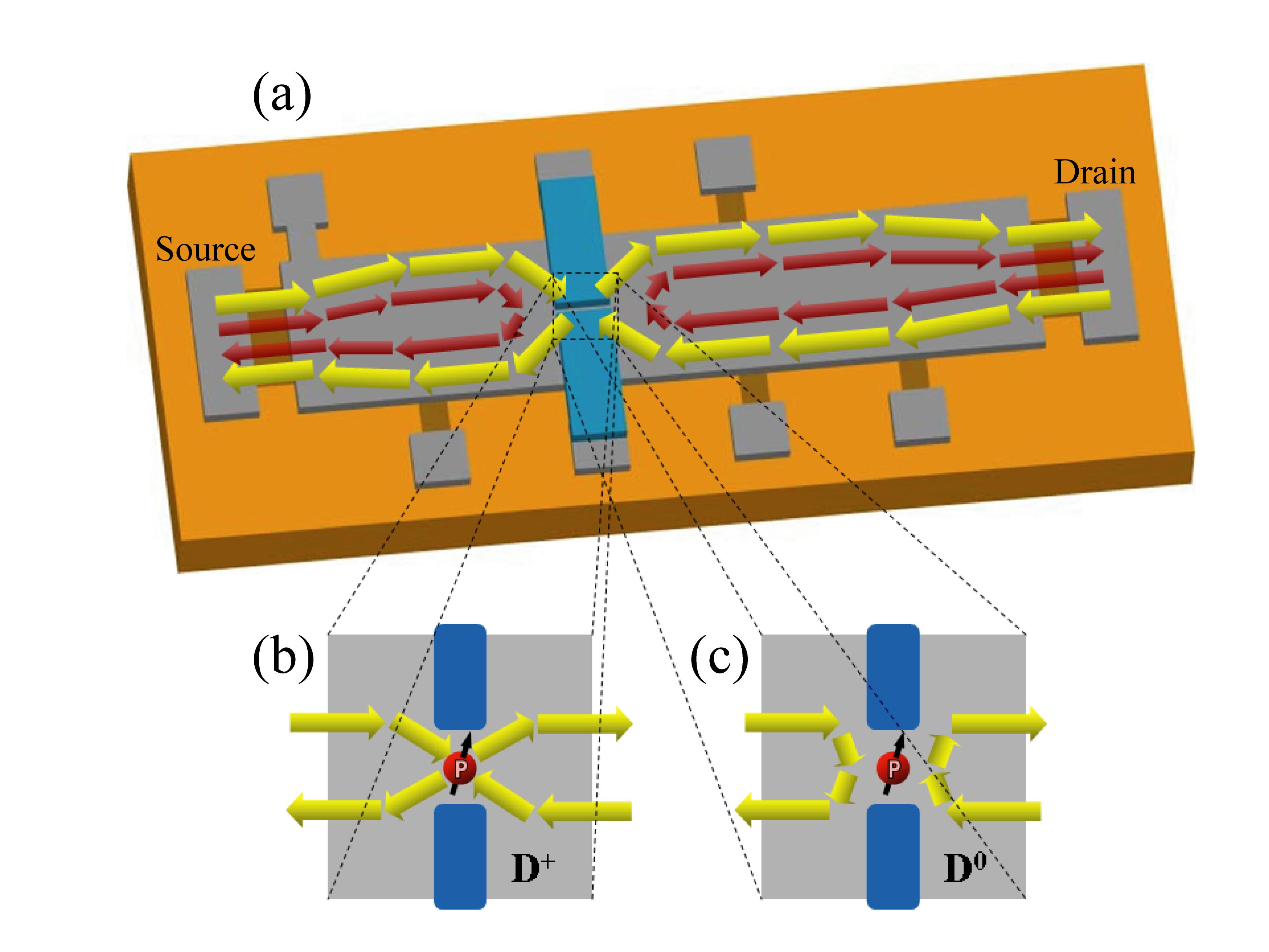} 
\end{center}
\caption{IQHE edge channels. (a) Only one occupied edge channel (yellow) tunnels into the QPC. (b) The edge channels are transmitted in the D$^+$ case, while (c) the edge channels reflect from the donor in the D$^0$ case.}
\label{iqheFig}
\end{figure}

The basis of the measurement device is a silicon metal
oxide semiconductor field-effect transistor (MOSFET)
Hall bar which exhibits the IQHE when placed in a static
perpendicular magnetic field at low temperatures. In the IQHE,
the transverse conductance across the Hall bar becomes
quantized and exhibits plateau regions as a function of
magnetic field, separated by step-function-like transitions. The
longitudinal conductance down the length of the Hall bar is
equal to zero during these plateaus and exhibits sharp peaks
during the transitions. This conductance quantization is a
result of the transformation of the momentum plane-wave
eigenstates of the 2DEG at zero field into edge channels at
nonzero field, localized at discrete distances from the
boundaries of the 2DEG (\fref{iqheFig}(a)). Both the distance between
edge channels and the width of the channels are determined by
the magnetic length $\tsc{l}{b} = \sqrt{\hbar/m^*\ts{\omega}{c}}$ where $\ts{\omega}{c}=eB/m^*$ is the electron cyclotron frequency, $m^*$ is the electron effective mass in silicon, and $B$ is the magnetic field~\cite{klitzingIQH}. The edge channels have energy spacing $\hbar
\ts{\omega}{c}$, which is significantly larger than the energy width of
each edge channel at low temperature determined by the Fermi-Dirac distribution. Consequently, the 2DEG conductance is quantized whenever the Fermi energy
falls between two edge channel energies, which is described as
having an integer filling factor $\nu$.

Our scheme takes advantage of the IQHE phenomena to improve the
robustness and decrease the noise in our measurement device. By operating
within the $\nu=1$ filling factor regime, only one edge channel is populated.
Due to the chirality of the edge channels, forward and backward
propagating channels lie spatially separated, on opposite sides of the Hall bar.
Since the next edge channel is separated
significantly in energy from the first ($\sim$ 0.7 meV at 2 T field, corresponding to $\sim$ 8 K), at low enough temperatures, electron scattering from one edge state to another by impurities or
other defects is negligible, and the measured conductance is insensitive to
these defects. Similarly, at the $\nu=1$ conductance plateaus, it is insensitive to magnetic field variations, further reducing noise.

While we do not want the measurement to be sensitive to most
defects in the device, we need the device to interact with one
particular defect, a P donor which is located just below the inversion layer in the center of the Hall bar. This donor could be implanted using single ion implantation through the optical aperture in the global gate~\cite{Weis08SingleIon}, or be placed via registered STM techniques~\cite{Ruess07STMIon}. To enable the coupling between the Hall bar and the donor, a QPC is located above the donor, insulated from the global gate by the oxide layer (\fref{deviceFig}). With an appropriate potential, the QPC restricts the edge channels and forces the two states with opposite wavevectors to slightly overlap, allowing scattering from one edge state to the other. Due to its proximity to the 2DEG, the donor
can scatter the edge channels within the QPC region, and the
scattering rate will be different when the donor is ionized
compared to when it is neutral. Edge state resonant scattering from a single impurity within a QPC has previously been observed and successfully simulated~\cite{kivelsonScattering}. However, this effect was only observed near a transition between plateau regions and only when the Fermi energy was resonant with the impurity state, and a change in scattering due to the ionization of the impurity was not studied.

\begin{figure}[t]
\begin{center}
\includegraphics[width=3.5in]{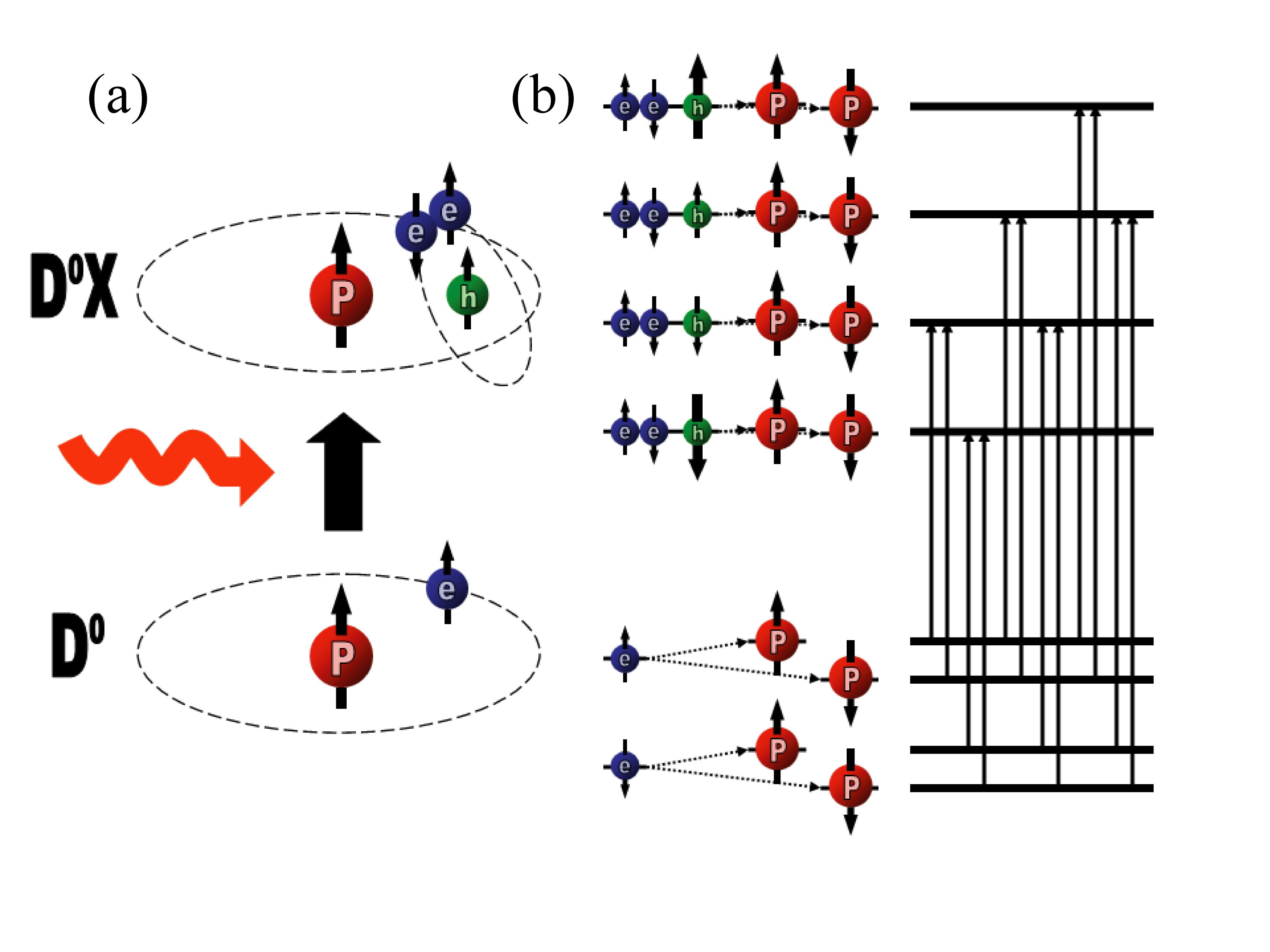} 
\end{center}
\caption{(a) Optical transition between the D$^0$ and D$^0$X states. The electrons form a single. (b) In a magnetic field, the D$^0$X state is split into the four Zeeman levels of the spin-$\frac{3}{2}$ hole, while the D$^0$ state is split into two electron Zeeman levels, each split again by the hyperfine coupling to the nuclear spin. Spin selection rules lead to 12 allowed optical transitions.}
\label{d0xFig}
\end{figure}

We can control the ionization state of the donor through the
use of a narrow linewidth CW laser tuned to the
D$^0\rightarrow$D$^0$X transition of the P donor (\fref{d0xFig}(a)). At
low temperatures, the neutral donor can bind a free exciton,
creating the 4-quasi-particle complex comprised of two electrons
coupled in a spin singlet, a hole, and the nucleus. This
metastable state can be resonantly excited with the laser, and
will then either decay optically or, three orders of magnitude more
often, non-radiatively via Auger recombination, leaving the
donor ionized. This decay mechanism is still relatively slow, leading to the very narrow linewidth of the optical transition. By optically pumping and
electrically detecting, we can take advantage of the narrow
optical transition while avoiding the very slow optical decay.

In a magnetic field, there are four nondegenerate D$^0$ states
- two electron Zeeman levels combined with hyperfine splitting
of each Zeeman level - and four nondegenerate D$^0$X states -
four hole Zeeman levels (the $p$-state-like hole does not
strongly couple to the nuclear spin) (\fref{d0xFig}(b)). Usually, the
hyperfine-split transitions are too broad to distinguish
because Si isotopes in the crystal have nuclear spin, which inhomogeneously broaden the transition linewidths due to bandgap fluctuations. However, we can remove
all isotopes other than the nuclear-spin-zero $^{28}$Si using isotopic purification, resulting in a linewidth that has been measured to be as narrow as 36 MHz~\cite{thewalt} with bulk photoluminescent excitation spectroscopy. More recently, a homogeneous linewidth of 2.4 MHz has been measured using spectral hole burning~\cite{thewaltLineWidth}, much narrower than the hyperfine splitting of 60 MHz.  Therefore, each of the 12 transitions
allowed by spin selection rules between the 8 states can be optically distinguished.

By tuning a laser to one of these transitions, we can selectively
ionize the donor only if it has the particular nuclear spin and electron spin corresponding to the ground state of that transition, otherwise leaving the donor neutral. If we excite a pair of transitions with the same electron spin state but opposite nuclear spin states (using either a pair of lasers or one alternating between these transitions), the ionization will only be dependent upon the nuclear spin state. This ionization will modify the transport of the edge
channels through the QPC until the donor recaptures another
electron from the 2DEG. The nuclear spin state will have negligible probability of flipping throughout hundreds of thousands of repetitions of this process~\cite{flsy04}, so the laser(s) can re-ionize the donor again, followed by electron recapture, and so on. By monitoring the transverse and
longitudinal conductivity across the Hall bar, we should
observe a random telegraph signal if the donor is in one particular
nuclear spin state, and no change in conductivity if the donor
is in the other spin state. If the change in conductivity is very large, a single-shot measurement would also be possible. Either method is a deterministic
quantum non-demolition measurement of the single donor nuclear spin.

\section{Device Physics and Simulation}
\label{DevicePhysicsSection}

In this section, we will discuss the important physical
effects that occur within our device, how we model these effects, and how to tune the device parameters to make the measurement feasible. Considerations include the effect of the oxide interface on the donor-bound electron, scattering of the edge channels
due to the donor, ionization and re-capture rates for the
donor, and optical linewidths and hyperfine splitting with
oxide-modified states.

\subsection{Donor Electron Ground State}
\label{EigenstateComputationSection}

The presence of the oxide layer next to the donor by necessity
will modify the donor electron state, and the distance from the
inversion layer to the donor is an important quantity. If the
donor is too close to the oxide, the inversion layer will
strip the electron from the donor. However, if the donor is too far
from the oxide, the donor potential cannot affect edge
channel scattering in the QPC. Ideally, the combination of the
donor and the 2DEG potentials will induce a donor-electron
ground state which is partially located at the donor position
and partially located within the inversion layer. Recent work
on single As donors near an oxide layer suggests that the desired
hybridization regime is obtainable~\cite{lansbergenDonor}.

In order to model the donor and 2DEG, we have numerically
calculated approximate eigenstates of the system's Hamiltonian. We
first construct an effective Hamiltonian for the system, making a number of simplifying assumptions. We assume a homogeneous, perfect crystal with a single effective mass $m^*$ (obtained by the geometric average of the effective mass along the three principal axes in Si), and at this time we ignore valley-orbit coupling and spin effects. Our Hamiltonian has the form
\be
H = \frac{1}{2m^*}\left(\frac{\hbar}{\tn{i}}\bi{\nabla} + e \bi{A}\right)^2 + V_{2\tc{deg}} + \tsc{V}{qpc} + \tsc{V}{d}.
\label{hamiltonian}
\ee
We take the vector potential to be $\bi{A}=Bx\hat{\bi{y}}$, and have used a field of 2 T in our calculations.

$V_{2\tc{deg}}$ defines the potential for the 2DEG inversion layer created by the global gate and the oxide. We take the origin to be located at the position of the donor and the positive $z$-axis to be perpendicular to and point towards the oxide interface. We approximate this potential as
\be
V_{2\tc{deg}} = \cases{
-2\gamma\frac{\exp{[(z-z_0)]/d]}}{1+\exp{[(z-z_0)/d]}} &for $z < z_0$ \\
\qquad\tsc{V}{b}\ &for $z > z_0$ \\
}
\ee
where $z_0$ is the distance from the donor to the
oxide, $d$ is the width of the inversion layer (taken to be 5 nm in our calculations), $\gamma$ is
the depth of the interface potential (taken to be 15 meV), and $\tsc{V}{b}$ is the energy difference between the Si band edge and the oxide band edge (3 eV).

$\tsc{V}{qpc}$ is the potential term for the QPC channel, which we
have modeled as a parabolic potential along the $x$-axis, which is perpendicular to the direction of the edge state propagation. We assume that the QPC is centered on the donor and that the potential is uniform along the length of the QPC channel in the $y$-direction, giving us
\be
\tsc{V}{qpc} = \frac{1}{2}m^*\tsc{\omega}{q}^2x^2.
\ee
Here, $\tsc{\omega}{q}$ defines the strength of the QPC
confinement, chosen to produce a potential with a shape of 0.26 $\mu$eV/nm$^2$ in our calculations.

$\tsc{V}{d}$ is the effective donor nucleus potential, which is
approximated as
\be
\ts{V}{d} = -\frac{\hbar^2}{m^*a^*}\frac{\beta}{\sqrt{r^2+r_\tn{s}^2}},
\label{Vdonor}
\ee
where $r^2=x^2+y^2+z^2$, $m^*$ is the effective mass
averaged over the three directions (0.33), $a^*$
is the effective Bohr radius in Si from effective mass theory (20 \AA) and $\beta=(\ts{E}{obs}/\ts{E}{eff mass})^{1/2} \simeq$ 1.26 is a correction to effective mass theory which gives the correct bulk binding energy \cite{siBook}. $\ts{r}{s}$ is a phenomenological screening distance which simulates inner-shell electron screening (taken to be 5 \AA).

We next use this effective Hamiltonian to calculate approximate eigenstates of the system, by constructing a set of basis states and diagonalizing the Hamiltonian in this basis. We expect that our eigenfunctions will be hybridized wavefunctions with a donor-electron-like component and an edge-channel-like component. We employ two separate orthonormal sets of basis states, one of approximate eigenstates for the donor electron and one of approximate
eigenstates for the 2DEG edge states, and orthogonalize the combination of basis states using the standard Gram-Schmidt orthonormalization procedure. By using the combined basis, our basis states will match the shape of our eigenstates well, aiding the numerical computations.

For the donor-electron-like basis states, we simply use normalized hydrogenic wavefunctions centered on the donor, notated as $\ket{\psi_{nlm}}$ with
\be
\psi_{nlm} = \tn{R}_{nl}(r)\tn{Y}_{lm}(\theta,\phi).
\ee
While these donor states are not exact eigenstates of $\tsc{V}{d}$, they are similar enough to form a good basis set for the donor-electron-like portion of the wavefunction. In the absence of an oxide, the donor electron energy is approximated by the energy of these states in a Coulomb potential,
\be
E_{nlm} = -\frac{\hbar^2\beta^2}{2m^*{a^*}^2}\frac{1}{n^2}.
\ee

For the edge-channel-like basis states, we use the normalized product of Hermite-Gaussian functions in the $x$-direction, momentum plane-wave functions in the $y$-direction, and Airy functions in the $z$-direction, notated as $\ket{\phi_{pqk}}$ where
\begin{eqnarray}
\phi_{pqk} = \tn{X}_{p}(x-x_k)\tn{Z}_{q}(z-z_0)\frac{1}{\sqrt{\ts{L}{c}}}\tn{e}^{\tn{i}ky} \label{channelState}
\\
\tn{X}_{p}(x-x_k) =
\frac{1}{\sqrt{2^p p!}} \frac{\exp\left[-\left(x-x_k\right)^2/2\sigma_x\right]}{\left(\sigma_x^2\pi\right)^{1/4}} \tn{H}_{p}\left[\frac{\left(x-x_k\right)}{\sigma_{x}}\right]
\\
\tn{Z}_{q}(z-z_0) = \frac{1}{\left|\tn{Ai}'(\alpha_q^0)\right|\sqrt{\sigma_{z}}} \tn{Ai}\left[\alpha_{q}^0-\frac{(z-z_0)}{\sigma_{z}}\right].
\end{eqnarray}
Here, $\alpha_{q}^0$ is the $q$th root of the Airy function and $\ts{L}{c}$ is the channel length, which defines the set of allowed wavevectors and is taken to be 300 nm. The widths $\sigma_{x}$ and $\sigma_{z}$ are given by
\begin{eqnarray}
\sigma_{x} = \sqrt{\frac{\hbar}{m^*\ts{\omega}{t}}}
\\
\sigma_{z} = \left(\frac{\hbar^2 d}{\gamma m^*}\right)^\frac{1}{3},
\end{eqnarray}
and $x_k$ indicates the displacement of the chiral edge states from the center of the QPC
\be
x_k = - \frac{\hbar k_y}{m^*}\frac{\ts{\omega}{c}}{\ts{\omega}{t}^2},
\ee
where $\ts{\omega}{t}=\sqrt{\tsc{\omega}{q}^2+\ts{\omega}{c}^2}$ is the combination of the QPC frequency $\tsc{\omega}{q}$ with the cyclotron frequency of the electron in the magnetic field $\ts{\omega}{c} = eB/m^*$.

\begin{figure}[b]
\begin{center}
\includegraphics[width=3.5in]{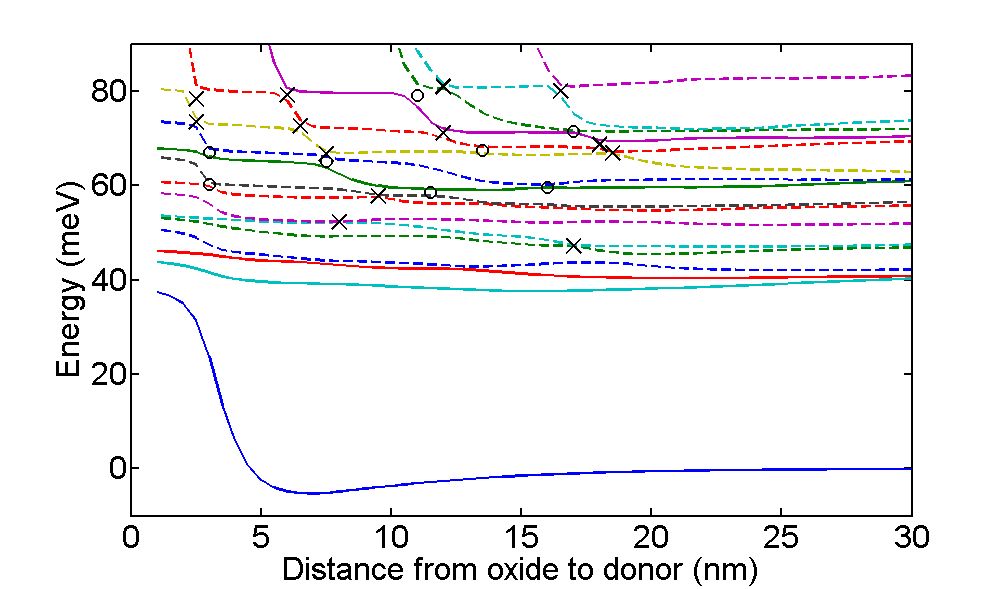} 
\end{center}
\caption{Energies of eigenstates as the distance between the donor and oxide is varied, measured with respect to the binding energy of P in bulk. Solid lines indicate donor-like states $\ket{\Psi_m}$ and dotted lines indicate 2DEG-like states $\ket{\Phi_n}$. O indicates anti-crossing, and X indicates crossing.}
\label{stateenergiesFig}
\end{figure}

The product of the Hermite-Gaussians in the $x$-direction and the momentum eigenstates in the $y$-direction are exact edge-state-like eigenstates of the combination of $\tsc{V}{qpc}$ and a magnetic field along the $z$-axis. The Airy functions are not exact eigenstates of $V_{2\tc{deg}}$, but are eigenstates of the triange potential $\tsc{V}{tri}$ where
\be
\tsc{V}{tri} = \cases{
-\gamma(1+(z-z_0)/2d) &for $z < z_0$ \\
\qquad\infty &for $z > z_0$ \\
},
\ee
which is approximately equal to $V_{2\tc{deg}}$ close
to the oxide~\cite{sternOxide}. As a result, the states are very similar to the eigenstates of the 2DEG potential. Therefore, this combination of the Hermite-Gaussian and plane-wave in the $xy$-plane and the Airy functions along the $z$-direction form a good basis set for the 2DEG edge-state-like portion of the wavefunction. In the absence of the donor, the energy of these states is approximated by their energy in the triangle potential $\tsc{V}{tri}$ and the QPC potential $\tsc{V}{qcp}$, giving us
\be
E_{pqk} = \hbar\ts{\omega}{t}\left(p+\frac{1}{2}\right) + \frac{\hbar^2k^2}{2m^*}\frac{\tsc{\omega}{q}^2}{\ts{\omega}{t}^2} -
\frac{\hbar^2}{2m^*\sigma_{z}^2}\alpha_{q}^0 - \gamma.
\label{channelEnergy}
\ee

\begin{figure}[b]
\begin{center}
\includegraphics[width=3.5in]{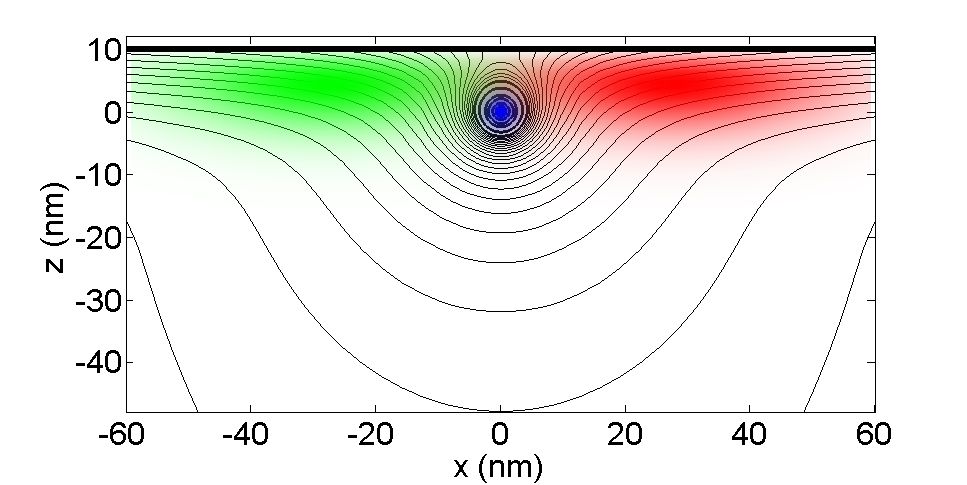} 
\end{center}
\caption{Contour plot of the potential energy in the $xy$ plane, for a donor that is 10 nm from the oxide. Plot also shows the amplitude of three eigenstate wavefunctions: the lowest donor-like state in blue, a forward-propagating edge state in red, and a backward-propagating edge state in green.}
\label{contourFig}
\end{figure}

Our two sets of basis states are not orthogonal, which is necessary for Hamiltonian diagonalization. Therefore, we choose a subset of basis functions from each type
of state and orthogonalize them using the standard Gram-Schmidt
orthonormalization procedure. After this, we can rewrite our
Hamiltonian in terms of these new basis states and diagonalize
it to find the hybridized eigenstates of the donor electron. Some of these states will be more donor-like, and we will label those states as $\ket{\Psi_m}$. The states that are more 2DEG-like will be labeled as $\ket{\Phi_n}$. \Fref{stateenergiesFig} shows the energies of a few of these eigenstates as a function of the distance between the donor and the oxide, and \fref{contourFig} shows the wavefunction in blue for the lowest energy donor-like state $\ket{\Psi_0}$ at a donor-oxide distance of 10 nm.

From these simulation results, we can draw a number of conclusions. We notice that the binding energy of the lowest D$^0$-like state is not significantly modified by the presence of the oxide unless the donor is less than 5 nm from the oxide. This is important for three reasons. First, the P donor will continue to bind an electron while close to the oxide and the 2DEG. Second, the fact that the electron binding energy changes only very slightly suggests that the donor will also bind an exciton in the presence of the oxide. Hayne's rule is a very successful empirical equation which says that for donors in bulk Si, the exciton binding energy is proportional to the electron binding energy, $\tsc{E}{x} \simeq \tsc{E}{d}/10$ ~\cite{haynesRule}. Since it is very challenging to calculate the binding energy of the complex four-quasi-particle D$^0$X state in the presence of an oxide layer, we will take Hayne's rule as the best approximation, and assume that the exciton binding energy of 5 meV will only vary in proportion to the electron binding energy. Finally, as we will discuss in \sref{opticalPropertiesSection}, we expect that the optical properties of the D$^0$X state such as the optical transition linewidth will not be significantly modified.

Analysis of the wavefunction of the D$^0$ lowest-energy eigenstate also shows that the wavefunction amplitude at the position of the donor is not significantly modified unless the donor is less than 5 nm from the oxide. This is important because it tells us that the hyperfine coupling between the donor electron and the nucleus should still be close to the bulk value, which is also discussed in \sref{opticalPropertiesSection}.

This simulation has given us an important lower-bound to the distance of the oxide from the donor. The next section will provide an upper-bound, at which point we'll discuss the tolerances of this distance, and what methods we have to tune our device in order to detect a donor at various depths.

\subsection{Edge Channel Scattering}\label{ChannelScatteringSection}

Outside of the QPC, edge channel states are spatially separated, and we assume there is negligible scattering from one state to another. However, within the
QPC, the tightly confined parabolic potential causes the edge
states to overlap, allowing mixing. By tuning the magnetic
field and the QPC voltage, we can allow only the lowest Landau level ($\nu=1$)
to tunnel through the QPC. Within the confinement of the QPC, the forward- and backward-propagating edge channels will overlap with the donor potential, where they will be scattered with some amplitude. This amplitude will change when the donor becomes ionized, which can be detected by monitoring the conductance through and reflection from the QPC.

The scattering potential in the ionized donor case is simply $\tsc{V}{d}$ from \eref{Vdonor}. For the neutral donor case, we must use the donor-bound-electron potential $\ts{V}{e}$ in addition to the donor potential $\tsc{V}{d}$. This potential describes the Coulomb interaction, including exchange terms, between the donor-bound-electron in the lowest energy state having wavefunction $\ket{\Psi_0}$ (as computed in \sref{EigenstateComputationSection}), and a second electron. This potential can be written as
\be
\fl\qquad \ts{V}{e} = \frac{\hbar^2\beta}{m^*a^*} \int\tn{d}\bi{x}\,\int\tn{d}\bi{x}'\, \left(\frac{\ket{\bi{x}}\braket{\Psi_0}{\bi{x}'}\braket{\bi{x}'}{\Psi_0}\bra{\bi{x}}}
{\left|\bi{x}-\bi{x}'\right|}
- \frac{\ket{\bi{x}}\braket{\bi{x}}{\Psi_0}\braket{\Psi_0}{\bi{x}'}\bra{\bi{x}'}}
{\left|\bi{x}-\bi{x}'\right|}\right).
\ee
For this computation, we have assumed the electrons are spin polarized.

To calculate the scattering amplitudes, we use the edge-channel basis states from \eref{channelState}. We assume that the occupied edge channel states are well represented by the lowest energy states, with $p = q = 0$ and $k$ determined by the Fermi energy. The scattering amplitude matrix elements for the scattering from a forward-propagating edge channel with wavevector $+k$ to a backward-propagating edge channel with the same energy and wavevector $-k$ are
\begin{eqnarray}
S_k^0 = \bra{\phi_{00-k}}\tsc{V}{d}+\ts{V}{e}\ket{\phi_{00+k}}
\label{ScatD0}
\\
S_k^+ = \bra{\phi_{00-k}}\tsc{V}{d}\ket{\phi_{00+k}}.
\label{ScatD+}
\end{eqnarray}

\begin{figure}[t]
\begin{center}
\includegraphics[width=3.5in]{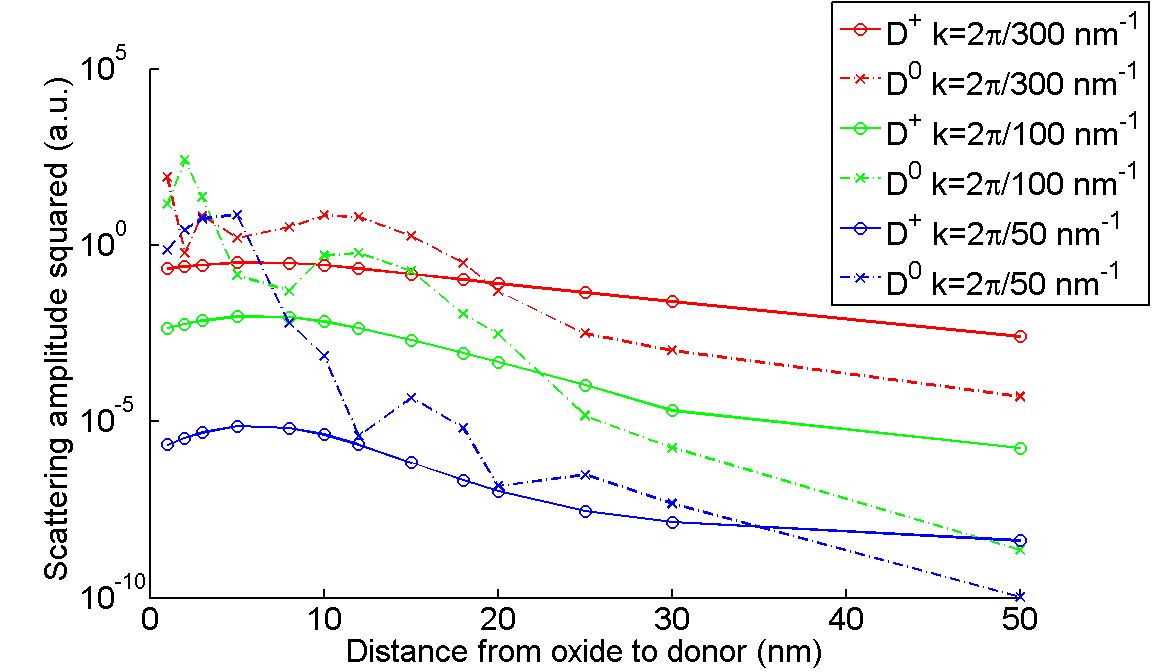} 
\end{center}
\caption{Amplitude squared of edge state scattering between edge states off the neutral donor within the QPC as a function of the distance between the oxide and the donor. Scattering for edge states with three different wavevector in both the ionized (solid line) and neutral (dotted line) donor cases is shown.}
\label{scatteringFig}
\end{figure}

The squares of these scattering amplitudes are shown for three different wavevectors in \fref{scatteringFig}. The difference between the squared matrix elements in the neutral case (dashed lines) and the ionized case (solid lines) shows the first-order change in scattering due to ionization of the donor, and gives strong indication that this change will be substantial for donor depths less than about 20 nm.

The fact that the interaction with the donor is relatively strong suggests that a significant portion of a conductivity quantum would be reflected by the presence of the donor. Furthermore, we see that the scattering amplitudes can be tuned by many orders of magnitude by adjusting the wavevector (or equivalently the Fermi energy), from negligible scattering with large $k$, to strong scattering with smaller $k$.

We can also gain some insight from the relationship between the scattering matrix elements in the two cases. First, we notice there are two interaction regimes. When the donor is close to the oxide, where the exchange interaction is strongest, the edge channels are scattered more by the presence of the donor electron than by the donor alone. When far from the oxide, where Coulomb interaction dominates over the exchange interaction, the edge channels are scattered more by the ionized donor than by the neutral donor.  In each regime, the scattering rates differ between the two cases by a few orders of magnitude, which suggests that there will be a large difference in conductivity, a necessity for the success of this measurement scheme. In order to achieve both strong scattering and a large difference in scattering between the ionized and neutral donor cases, we will likely work in the exchange-interaction-dominated region because the absolute strength of the scattering is much stronger there. The switchover between regimes happens when the scattering from the neutral and ionized donor is equivalent, around 20 nm (this exact value varies with wavevector). This sets an upper limit on the depth of the donor where the measurement could still be successful.

When scattering rates are small (i.e. when the donor depth or the wavevector $k$ is large) or the interaction time is very short, these scattering matrix elements can be combined with the Landauer-Buttiker formalism to estimate conductivity~\cite{fisherLeeCond}. This formalism requires the calculation of a reflection coefficient $R_k$, which indicates the fraction of the edge-channel with a particular wavevector $k$ that scatters from the QPC. We estimate this by taking the scattering rate from Fermi's golden rule and multiplying by the interaction time $\tsc{t}{i}$, which produces
\be
R_{k} = \frac{2\pi\tsc{t}{i}}{\hbar}\rho(E_k) \left|S_k^{+,0}\right|^2.
\ee
Here, $\rho(E_k)$ is the 1D density of states of the edge-channels with wavevector $k$ confined to the channel length $\ts{L}{c}$ from \eref{channelState},
\be
\rho(E_k) = \frac{1}{\hbar}\sqrt{\frac{m^*}{2E_k}} \left(\frac{\ts{L}{c}}{2\pi}\right)
\left(\frac{\ts{\omega}{t}}{\tsc{\omega}{q}}\right),
\ee
and $E_k$ is the energy component of the edge-channel states in \eref{channelEnergy} which is determined by $k$,
\be
E_k = \frac{\hbar^2k^2}{2m^*}\left(\frac{\tsc{\omega}{q}}{\ts{\omega}{t}}\right)^2.
\ee
The interaction time $\tsc{t}{i}$ is estimated by dividing the interaction length $\tsc{L}{i}$ by the group velocity of the edge-channel state $\ts{v}{g}$, which is also obtained from \eref{channelEnergy}. From this approximation, we find
\be
\tsc{t}{i} = \frac{\tsc{L}{i}}{\ts{v}{g}} = \tsc{L}{i}\sqrt{\frac{m^*}{2E_k}}\left(\frac{\ts{\omega}{t}}{\tsc{\omega}{q}}\right).
\label{intTime}
\ee
where we have taken $\tsc{L}{i}$ to be twice the effective bohr radius ($\simeq$ 40 \AA) for our calculations.

This approximation allows an estimation of $R_k$ for both the neutral and ionized donor cases; for example, at a relatively large wavevector $k = 2\pi/50~\tn{nm}^{-1}$ and a donor-oxide distance of 10 nm, we estimate that $R_k$ changes from 0.03 to 0.0002 upon ionization of the donor. The corresponding transverse conductance values can be obtained from the Landauer-Buttiker formalism,
\be
G_{xy}^{0,+} = \frac{e^2}{h}\frac{\tsc{L}{c}\tsc{L}{i}k^2}{4E_k\,^2} \left|S_k^{0,+}\right|^2,
\ee
and are 1 M$\Omega^{-1}$ for the neutral donor and 0.008 M$\Omega^{-1}$ for the ionized donor, a difference which can be measured by current equipment. For smaller $k$ values, the scattering matrix elements are substantially higher, in many cases causing the first-order reflection coefficient to exceed unity. This indicates the need for inclusion of higher-order interference terms within the formalism. For these $k$ values, even larger conductance changes are expected.

In \sref{EigenstateComputationSection}, we gave an effective lower-limit to the depth of the donor of 5 nm, and in this section we found an upper limit of around 20 nm. Note that this is only for a single set of device parameters, and in an actual experiment we will have a range of methods for tuning the interaction. Using the global gate, we will be able to tune the depth and width of the 2DEG. By varying the source-gate voltage, we can tune the Fermi energy of the current-carrying wavevector state, and by tuning the QPC voltage, we can ensure only a single edge channel is mixed within the QPC. This gives us a very large parameter space which should encompass the desired interaction regime for a broad range of donor depths.

\subsection{Ionization and Recapture}
\label{IonizationSection}

The ionization rate of laser excitation and subsequent Auger
recombination is limited by the lifetime of the D$^0$X state,
which is 272 ns in bulk and has been shown experimentally to vary with binding energy $\tsc{E}{d}$ as $\tau \propto \tsc{E}{d}^{-3.9}$~\cite{schmid77}. If our donor is in the optimal range of 5 to 20 nm, the donor binding energy varies only a small percentage from the bulk value, and so we expect that the excited state lifetime when the donor is near the oxide should not differ significantly from the lifetime in bulk.

After the donor is ionized, an edge state electron can be
scattered into the donor-bound-electron state, re-neutralizing
the donor. This recapture is important for the measurement
scheme, as it allows the laser to re-ionize the donor to continue the cycle,
producing a random telegraph signal. In particular, the recapture time is crucial; if it is much faster than a few nanoseconds, the ionized state will come and go too fast for detection. If it is too slow, then the time required to repeat the cycle maybe become prohibitively long. Unfortunately, the strongly-coupled, many-body nature of the relevant interactions makes the type of perturbative approaches we have used so far ineffective to estimate this recapture rate. Experiments in optical spectroscopy in the presence of above-band carriers~\cite{thewalt}, however, suggest that this recapture rate could be made to fall into a range reasonable for an effective measurement with appropriate tuning of the bias current.

Any ionization process which competes with the Auger decay will manifest itself as noise in this measurement. Two such ionization processes are thermal ionization and field impact ionization due to the electric field near the oxide. Since all of the experiments will be conducted below 4 K, the donor binding energy of 45 meV is much larger than $\tsc{k}{b}T$ ($\simeq$ 0.3 meV), making thermal ionization negligible.

Field impact ionization of the donor electron is a concern, as that can occur at fields of 400 V/cm~\cite{wemanBEField1}, which is about half of the field felt by a donor 20 nm from the oxide. However, the field creating the 2DEG is only felt in the near vicinity of the oxide and does not have enough distance to give a free carrier the energy necessary to ionize the donor. Furthermore, most of the free carriers in bulk are frozen out at low temperatures.

Field impact dissociation of the exciton is a much greater concern, as this can occur at 50 V/cm~\cite{wemanBEField1}. However, instead of ionizing the donor, this effect would prevent the donor from being ionized. As with impact ionization of the donor electron, the field range is very small and carriers will be mostly frozen out at low temperature, greatly reducing the rate of this effect. Future devices could be optimized to reduce the 2DEG field strength in order to limit this effect.

The electric field itself can also pull an electron or exciton off the donor, but that is a negligible effect below 35 kV/cm~\cite{martinsD0Efield} for electrons and below 5 kV/cm~\cite{blosseyEXs} for excitons, and would be negligible for appropriate donor depths.

\subsection{Optical Transition}
\label{opticalPropertiesSection}

The narrow linewidth of the D$^0$ $\rightarrow$ D$^0$X
transition is a crucial requirement for the success of our proposed experiment.
In order to selectively ionize the donor in a particular spin
state, the linewidth must be smaller than the hyperfine
interaction energy. Fortunately, a bulk sample within a strong
magnetic field has a hyperfine splitting of 60 MHz and the best
measurement of the donor-bound-exciton transition homogeneous linewidth is 2.4 MHz ~\cite{thewaltLineWidth}. However, this linewidth could increase in the presence of the oxide due to excited-state-lifetime changes or additional oxide defects.

Near the oxide, we do not expect lifetime shortening to increase the linewidth beyond 2.4 MHz, since that value is already four times the lifetime-limited linewidth, and as discussed in \sref{IonizationSection}, this lifetime should not be significantly modified by the presence of the oxide. Oxide defects, however, could broaden the linewidth due to mechanisms such as spectral diffusion of oxide defects. As long as these defects do not shift the donor states excessively fast or strong, the linewidth should be similar to the bulk case.

The hyperfine splitting will change slightly due to the oxide since the splitting is proportional to the square of the electron wavefunction amplitude at the position of the donor nucleus~\cite{siBook}, and that amplitude varies with the donor-oxide distance because the
donor-bound electron wavefunction shifts slightly into the 2DEG when the donor is close to the oxide. However, from analyzing the hybridized donor-bound-electron wavefunction we
calculated in \sref{EigenstateComputationSection}, this amplitude does not change by more than 4\% for a donor further than 5 nm from the oxide, resulting in less than a 10\% change in the hyperfine splitting.

Central-cell corrections to the binding energy of the donor-bound-electron should also be negligible since the amplitude of the electron wavefunction throughout the central-cell region similarly does not change significantly. Furthermore, any binding energy shifts due to central-cell corrections or other effects, such as the DC Stark effect, that do not cause broadening can easily be compensated for by tuning the laser. For this reason, and because the hyperfine splitting should remain significantly larger than the bulk homogeneous linewidth, selective ionization of individual donor states should still be possible in the presence of the oxide, even with moderate broadening due to nearby defects.

\subsection{Photoconductivity}

A major concern for photoconductivity measurements in Si and other semiconductors is free electron creation from illuminated metallic leads. Photocurrent originating from these metallic leads will greatly increase background noise. Due to the rather long wavelength of the excitation photons, even a diffraction limited laser spot incident on the device would likely cause a significant amount of background photocurrent. For this reason, the metallic global gate will also perform duty as a beam block to protect the metallic leads from the laser illumination in addition to being used to tune the 2DEG (see \fref{deviceFig}). A small aperture will be created in the global gate above the position of the donor to protect the leads in proximity to the donor. The global gate and the QPC gates will be illuminated, but they are electrically isolated from the source, drain, and measurement electrodes, and will not directly produce noise. An antireflection coating on the back of the sample may also be required to reduce reflection from the back of the sample.

\section{Conclusion}

In summary, by combining optical pumping and electronic
detection methods together into one system, we are able to take
advantage of the benefits of each method in order to overcome
the difficulties presented by semiconductor systems and create
a realistic measurement device. This measurement scheme is a
deterministic non-demolition measurement of the
nuclear spin of a single $^{31}$P in Si. If single-shot measurement is achieved, then by optically pumping on a pair of transitions beginning in the same electron spin state but opposite nuclear spin state we can also a perform deterministic measurement of the donor-bound-electron spin, but in this case it is a destructive measurement.

A number of other factors will contribute to this measurement scheme. In particular, we have not included the critical details of the Si band structure. The different valley-orbit states make the scattering problem more complicated, reduce the symmetry, and introduce constraints on the donor placement. Further, the inversion layer is treated here as an empty, triangle-like potential well, but in reality there is a bath of electrons in this potential, and the many-body effects of screening and spin-spin scattering will play an important role inside the QPC. The present discussion is intended to introduce the principle of our measurement scheme; more detailed calculations including these important effects are in progress.

\ack

DS was supported by Matsushita through the Center for Integrated Systems. This work was supported by the University of Tokyo Special Coordination Funds for Promoting Science and Technology, NICT, and MEXT.

\section*{References}


\end{document}